\documentclass[twocolumn,showpacs,superscriptaddress,10pt,pra]{revtex4-1} 
\usepackage{amsmath,amssymb,graphicx,physics,float}

\begin{document}

\title{Probing local density of states near the diffraction limit using nanowaveguide coupled cathode luminescence}
\author{Yoshinori Uemura}
\author{Masaru Irita}
\author{Yoshikazu Homma}
\author{Mark Sadgrove$^*$}

\affiliation{Department of Physics, Faculty of Science, Tokyo University of Science, 1-3 Kagurazaka, Shinjuku-ku, Tokyo 162-8601, Japan}
\email{mark.sadgrove@rs.tus.ac.jp}

\begin{abstract}
The photonic local density of states (PLDOS) determines the light matter interaction strength in nanophotonic devices.
For standard dielectric devices, the PLDOS is fundamentally limited by diffraction, but its precise dependence on the size parameter $s$ of a device can be non-trivial. Here, we measure the PLDOS dependence on the size parameter in a waveguide  using a new technique - nanowaveguide coupled cathode luminescence (CL). We observe that  depending on the position within the waveguide cross-section, the effective diffraction limit of the PLDOS varies, and the PLDOS peak shape changes. Our results are of fundamental importance for optimizing coupling to nanophotonic devices, and also open new avenues for spectroscopy based on evanescently coupled CL.
\end{abstract}
\maketitle
\section{Introduction} 
The rate of decay of an emitter into a given optical mode is governed by Fermi's golden rule, and is proportional
to the photonic local density of states (PLDOS) $\rho$ associated with that mode. A fundamental limit on $\rho$ for 
nanophotonic devices is the diffraction limit which places a lower bound on the mode size of $\sim\lambda/2$ 
in a given dimension~\cite{novotny2012principles}. Dielectric devices with a characteristic size less than this have sub-optimal PLDOS due to redistribution of mode amplitude into the evanescent region - i.e. a loss of mode confinement.
An operational definition of the diffraction limit for nanodevices is, therefore, the size at which the PLDOS is maximized.

An important class of diffraction limited nano devices is that of nanowaveguides, which are used in fields ranging from quantum optics~\cite{aharonovich2016solid} and optomechanics~\cite{khanaliloo2015single} through to particle manipulation~\cite{yang2009optical}. For certain nanowaveguide types, systematic measurement of the photonic local density of states via cathode luminescence (CL) spectroscopy~\cite{de2010optical,polman2019electron,brenny2016near} has been achieved via leaky modes. In this remarkable technique, depicted in Fig.~\ref{fig:principle}(a), electrons incident on a device induce luminescence, offering essentially tomographic PLDOS reconstruction due to the point-dipole-like excitation provided by the electron beam~\cite{atre2015nanoscale,sapienza2012deep,horl2017tomographic}. However, because luminescence is collected in the far-field, the PLDOS of true waveguide modes (which by definition do not couple to radiation modes) cannot be measured in general. Furthermore, although it is well known that an optimal diameter exists for coupling to nanowaveguides~\cite{le2005spontaneous}, no systematic measurement of the diffraction limited behavior of waveguide PLDOS has ever been performed to the best of our knowledge.

Here, we detect CL emitted into a the fundamental mode of a nanowaveguide (optical fiber taper) as depicted in Fig.~\ref{fig:principle}(b). We use this new technique to characterize hitherto unmeasured aspects of the waveguide mode PLDOS. In particular, we measure the PLDOS dependence on the waveguide size parameter $s$ (defined below) around the diffraction limit.  Using different electron energies, we probe the PLDOS i) close  to the waveguide surface, where the near-field character of the mode is strong, and ii) nearer to the waveguide center where the mode has a standard transverse wave character. These two regimes are shown to exhibit different dependence on the size parameter, and in particular a different effective diffraction limit. These results shed light on a fundamental characteristic of nanowaveguides, and illuminate the subtle nature of 
the widely used diffraction limit concept for nanohotonic devices. Furthermore, the new method of waveguide-coupled CL promises a novel way to create fiber coupled electrically driven photon sources and probe previously inaccessible characteristics of optical near-fields using the CL technique.

\section{Principle and methods}
\begin{figure*}
\centering
\includegraphics[width=0.65\linewidth]{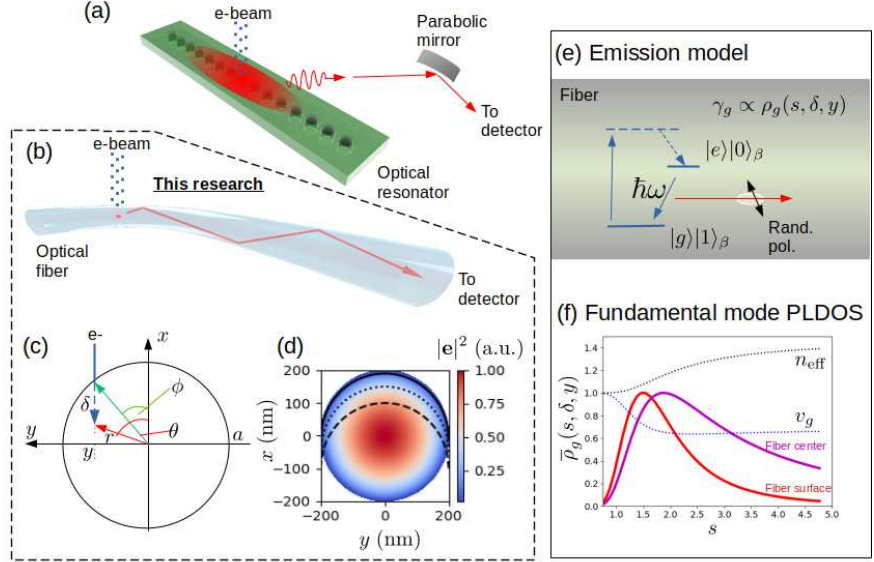}
\caption{\label{fig:principle}Principle of the experiment. (a) Example of a standard cathode luminescence spectroscopy experiment. A resonant mode leaks photons which reach a detector in the far field. (b) Concept of the present work.  Electrons are incident on a vacuum clad optical fiber of radius $a$ and CL is detected through the guided mode itself. (c) Electrons incident at a point $(a,\phi)$ penetrate a distance $\delta$ into the fiber to point $(r,\theta)$ and induce cathode luminescence 
which couples directly to the fiber fundamental mode. (d) Intensity $|\mathbf{e}|^2$ of a circularly-polarized fundamental (HE$_{11}$) mode of the fiber with curves showing electron stopping position for $\delta = 10$ nm (solid line), $\delta = 50$ nm (dotted line) and $\delta = 100$ nm (dashed line) 
(e) Emission model. The energetic electron is assumed to excite an emitter within the fiber silica matrix to a high energy level which then decays by non-radiative processes before emitting a randomly polarized photon into the fiber fundamental mode with propagation constant $\beta$ at a center wavelength near 659 nm. (f) The thick red (magenta) line shows the normalized photonic local density of states $\overline{\rho}_g$ at the fiber surface (center) nm as a function of the size parameter $s$.  Also shown are $v_g/c$ (dotted blue line), and the effective refractive index of the mode $n_{\rm eff}$ (dotted black line).}
\end{figure*}
The principle of our experiment is shown in Figs.~\ref{fig:principle}(b) and (c). 
Electrons from a scanning electron microscope (SEM) penetrate a vacuum clad silica fiber (core refractive index $n_{\rm co}=1.46$) of radius $a$ (200 nm $\leq a \leq$ 1 $\mu$m) to a depth $\delta$ which depends on the electron energy. The electrons induce luminescence in the silica, a portion of which couples directly to the fiber fundamental modes with an intensity that depends on the photonic local density of states of the modes. As shown in Fig.~\ref{fig:principle}(c), for a given value of $\delta$ and a position $y$ along the fiber cross section, the radial position $r$ and angle $\theta$ of the electron stopping position can be defined, with $\phi = \sin^{-1}(y/a)$, $r=\sqrt{y^2+(a\cos\phi-\delta)^2}$ and $\theta = \pi/2 - \cos^{-1}(y/r)$. In Fig.~\ref{fig:principle}(d), the so-parameterized stopping point of the electrons as a function of $y$ is overlaid on the profile of a fundamental fiber mode for the case where $a=200$ nm, and the CL wavelength is $659$ nm for three different values of $\delta$. 

As shown in Fig.~\ref{fig:principle}(e), we assume that the measured light is from incoherent CL~\cite{de2010optical} which is produced in an effective off-resonant excitation process in which unpaired oxygen defect centers in the silica~\cite{sigel1981photoluminescence} are excited to a high energy level which decays non-radiatively before a final radiative transition produces randomly polarized luminescence with a phonon-broadened spectrum. The emission is assumed to occur at the point in the material where the electron comes to a stop, i.e., a distance $\delta$ from the fiber surface. (In fact, the process is more complicated: a cascade of secondary electrons is also created after the primary electron enters the material, and CL can originate from these electrons too. For the 0.5 keV energy used predominantly in this work, this cascade region is approximately 10 nm in diameter. We treat this behavior phenomenologically by treating the electron beam as having a Gaussian distribution of a similar width and convolving this distribution with the PLDOS.)

Assuming a single mode fiber, the coupled intensity of the CL is proportional to the decay rate $\gamma_g$ into the fundamental  fiber modes at the position $\mathbf{r}_0$ in the fiber where CL is generated. 
In general we may write this relation as~\cite{novotny2012principles,sondergaard2001general}
$\gamma_g = \frac{2\mu_0\omega_0^2}{\hbar}{\rm Im}[\mathbf{p}\cdot \mathbf{G}_T(\mathbf{r}_0,\mathbf{r}_0,\omega_0)\cdot\mathbf{p}]$,
where $\omega_0$ is the transition resonant frequency, $\mathbf{p}$ is the dipole moment, and $\mathbf{G}_T$ is the guided mode transverse Green tensor.  The imaginary part of the Green tensor may be evaluated~\cite{le2018force,sondergaard2001general} yielding
${\rm Im}[\mathbf{G}^T(\mathbf{r}_0,\mathbf{r}_0,\omega_0)]=\frac{c^2\mathbf{e}(\mathbf{r}_0)\mathbf{e}^*(\mathbf{r}_0)}{4v_g\omega_0}$.
Here, $v_g$ is the mode group velocity and  $\mathbf{e}(\mathbf{r}_0)$ is taken to be the normalized mode function of the positive propagating,  left hand circular polarized HE$_{11}$ fundamental mode of the fiber. The mode function is normalized according to the condition $1 = \int d^2rn(r)^2|\mathbf{e}(\mathbf{r_0})|^2$, where the integral is taken over a plane perpendicular to the fiber axis.  The product of mode functions is interpreted as a dyad. Details of the mode functions are given in the Appendix. In our present study, the wavelength of the modes is fixed at $\lambda = 659$ nm, and the value that the mode function takes depends on the fiber radius $a$, at the radial position $\mathbf{r}_0(y,\delta)$.
Note that the quantity $|\mathbf{e}(\mathbf{r}_0)|^2$ has units m$^{-2}$ and may be considered to be a dimensionless- energy flux. This should be compared to the usual energy density associated with three dimensionally confined resonant modes.

By circular symmetry, a randomly polarized dipole couples with the same strength to either of the two orthogonally polarized
fundamental modes. We may average over dipole polarization to produce the photonic local density of states associated with the fundamental modes~\cite{novotny2012principles}
\begin{equation}
\rho_g(s,\mathbf{r}) = \frac{2}{3}\frac{6\omega_0}{\pi c^2}{\rm Im}[{\rm Tr} [\mathbf{G}(\mathbf{r}_0,\mathbf{r}_0,\omega_0)]]=\frac{|\mathbf{e}(s,\mathbf{r}_0)|^2}{v_g},
\label{eq:ldos}
\end{equation}
where the factor of $1/3$ arises from the average over dipole orientations, and the factor of 2 arises due to the two possible orthogonal polarizations of the fundamental mode.

Finally, we see that 
\begin{equation}
\overline{\gamma}_g =\frac{\pi\omega_0}{3\hbar\epsilon_0}p^2\rho_g(s,\mathbf{r}),
\end{equation}
where $\overline{\gamma}_g$ is the decay rate into the fundamental modes averaged over polarization, and the dipole moment strength is assumed to be $p=|\mathbf{p}|$ in any direction. Note that for a given $s$, $\rho_g$ contains all the dependence of $\overline{\gamma}_g$ on the fiber mode behavior. Our experimental measurements are of photon count rates through the fiber over some time $\Delta t$. It may be seen that such measurements are proportional to $\overline{\gamma}_g\Delta t\propto\rho_g$. In practice, we normalize both our measurements and the theoretical predictions for $\rho_g$ so that their maxima are equal to unity before comparing them. We denote the so-normalized value of the PLDOS by $\overline{\rho}_g$.

Because Maxwell's equations are scale free, the functional dependence of the local density of states on the waveguide transverse dimension $a$ or the wavelength $\lambda$ are most generally expressed using the dimensionless size parameter $s=k a = (c / \omega_0) a$, where $k=2\pi / \lambda$. By using a tapered fiber, we allow the measurement of the PLDOS as a function of $s$ for fixed $\lambda$ and variable $a$.

The thick red line in Fig.~\ref{fig:principle}(f) shows the normalized local density of states as a function of $s$ just inside the fiber surface. The thick magenta line shows the same calculation made at the fiber center. 
Also shown are the scaled group velocity of the fundamental mode $v_g/c$ (dotted blue line) and the effective refractive index $n_{\rm eff}$ for the fundamental mode (dotted black line). It may be seen that peak region of the PLDOS is associated with the transition of $v_g$ from the bulk silica value of $v_g\approx c/1.45$, to $v_g\approx c$ as the fiber mode is dominated by its evanescent component. 
Note that the maximum value of the unscaled PLDOS at the fiber center is almost three times larger than that just inside the fiber surface. Because the present experiment does not allow us to cleanly measure the relative amplitude of the PLDOS at these two different radial positions, we use the normalized PLDOS and focus on the differences seen in the peak position and peak width.

The most notable aspect of the PLDOS curves for different radial positions is that the peak value occurs at a different value of $s$. In this sense, the effective diffraction limit of $s$ is different depending on where in the fiber cross-section it is measured. This is a generic feature of waveguides (i.e. not just fibers) and occurs due to the behavior of the mode function $|\mathbf{e}(\mathbf{r})|=A(s)F(s,r)$, where $A(s)$ is a normalization factor depending only on the size parameter, and $F(s,r)$ is in general a decreasing function of the radial distance $r$ from the fiber center. Broadly speaking, $A(s)$ sets the intensity scale at a given value of $s$ for a fixed optical power,
and thus has a peaked structure which gives rise to the diffraction limit. $F(s,r)$ can generally be written in the form $F(u r/a)$,
where $u=a\sqrt{n_{\rm co}^2k^2-\beta^2}$ is a dimensionless wavenumber which increases monotonically with the waveguide size parameter $s$. As $r/a$ increases, the fall-off in $F$ as a function of $u$ becomes steeper, leading to the peak of the PLDOS occuring at lower $s$. This is also the reason for the the narrower width of the PLDOS peak when $r=a$ compared with $r=0$. More details are given in the supplementary material. In this sense, despite being polarization averaged, the PLDOS near the diffraction limit contains information about the near-field nature of the mode, which is transverse near the fiber center but vectorial in nature at the fiber surface. 

Experimentally, we detect the intensity in the fiber modes by passing a single mode fiber which is adiabatically connected to the fiber taper out of the SEM vacuum via a feedthrough. The fiber can be connected to a spectrum analyzer or a modified Hanbury-Brown-Twiss setup which allows measurement of both polarization and the intensity correlation function $g^{(2)}$. In experiments, we used electron energies of 0.5 keV in a spot excitation configuration, and 2 keV in a sweep excitation configuration. CL emitted into the fiber taper passed through a 630 nm cutoff single mode fiber to ensure that only light in the fundamental modes was collected. Further details of the experiment are given in the Appendix.

\section{Results}
\begin{figure}
\centering
\includegraphics[width=\linewidth]{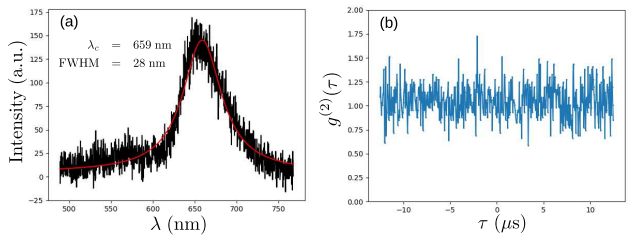}
\caption{\label{fig:results1}(a) Measured spectrum of the fiber coupled CL. (b) Measured second-order correlation function $g^{(2)}(\tau)$ for a time difference $\tau$ between detection events.  }
\end{figure}

We now turn to our experimental results. First, we look at general properties of the the fiber coupled cathode luminescence.
The CL spectrum measured through the guided modes is shown in Fig.~\ref{fig:results1}(a). A Lorentzian curve was fitted to the data and, as indicated, the center wavelength was found to be 659 nm and the full width at half maximum (FWHM) was found to be 28 nm. This spectrum is similar to that seen in silica fibers due to radiation induced defects, or the fiber drawing process itself~\cite{sigel1981photoluminescence}. The luminescence has been attributed to unpaired oxygen atoms in the silica matrix.

We also checked the polarization at the fiber output by rotating both a half waveplate and a quarter waveplate before the light entered a polarizing beam splitter, and measuring the output at both ports. For both waveplates, we saw variations in intensity of about $\pm 5\%$ of the mean value, suggesting nearly perfect random polarization. 

Because little is known about the density of defects in silica which produce the observed cathode luminescence, we also measured the count coincidence rate of the CL through the guided modes. The normalized coincidence signal corresponds to the second order correlation function $g^{(2)}(\tau) = \langle n(t)n(t+\tau)\rangle / (\langle n(t)\rangle\langle n(t+\tau)\rangle)$, where $n$ denotes photon counts, the coincidence delay is given by $\tau$, and $\langle \cdot \rangle$ denotes a time average. For a single or few emitters, an anti-bunching dip in the coincidence rate is expected at $\tau=0$. As seen in Fig.~\ref{fig:results1}(b), the measured correlation function shows no sign of antibunching and is consistent with a relatively large number of independent photon emitters within the excitation volume.

\begin{figure*}
\centering
\includegraphics[width=0.65\linewidth]{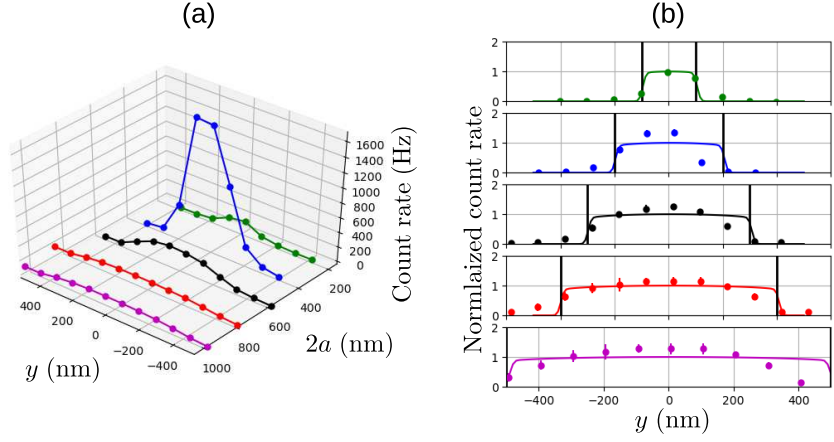}
\caption{\label{fig:results2} Spot scans perpendicular to the optical fiber axis for electron energies of 0.5 keV. (a) Shows unnormalized data (discrete points)
for five different fiber diameters with lines connecting points to guide the eye. (b) Shows the same data normlized and fitted by $\overline{\rho}_g(a,\delta,y)$ convolved with a Gaussian beam profile. From top to bottom, the data shown is for $2a=200$, 400, 600, 800, and 1000 nm. Theoretical curves for $\delta=10$ nm are shown for each case.}
\end{figure*}

Next, we consider scans made of the fiber over its cross section for fiber diameters between 200 and 1000 nm. Fig.~\ref{fig:results2}(a) shows raw count rates (discrete points) joined by lines to guide the eye.
It is notable that a large peak is observed at $2a=400$ nm relative to the other diameters. This is due to the 
increased mode confinement at this diameter.
Fig.~\ref{fig:results2}(b) shows the same experimental results normalized to allow easier comparison.
In each case, curves showing values of $\overline{\rho}_g(a,\delta,y)$ for $\delta=10$ nm convolved with a Gaussian profile with a standard deviation of 10 nm to account for the broad electron cascade process inside the silica. For these curves, we fitted the value of the amplitude and center position to the data. The fiber diameter was set to its experimentally measured value in the theory. Note that the colors of the points and curves correspond to the data shown in the same color in Fig.~\ref{fig:results2}(a). Error bars show $\pm$1 standard deviation over ten intensity measurements.  

The data show that the CL intensity varies only slowly across the fiber cross section.
This is expected considering the circular symmetry of the coupling, i.e., a randomly polarized emitter should couple with the same strength to the fundamental modes at any position within the fiber that is a constant radial distance from its center. However, due to the stopping position on the $x$ axis being dependent on $y$, the distance from the fiber center at which CL occurs changes with the change becoming larger as the penetration depth increases.

\begin{figure}
\centering
\includegraphics[width=0.8\linewidth]{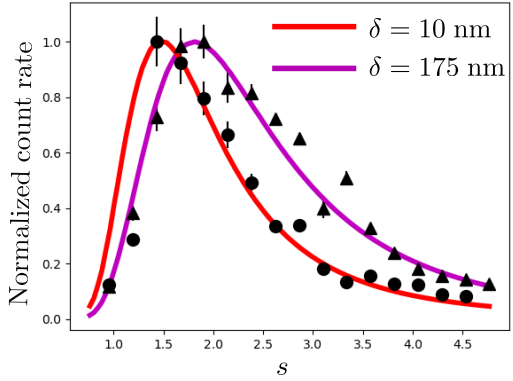}
\caption{\label{fig:results3} Measurement of relative PLDOS as a function of diameter. Circles show measurements made using a stationary electron beam of energy 0.5 keV at the fiber center. The measurements shown are the averaged raw data, with error bars showing the standard deviation over ten separate measurements. The red curve shows $\overline{\rho}_g(a, \delta=10\;{\rm nm}, y=0)$. Triangles show similar measurements, but for a beam energy of 2.0 keV, which corresponds to $\delta=175$ nm. The theoretical value of $\overline{\rho}$ in this case is shown by the magenta curve. }
\end{figure}
Finally, we measured the waveguide coupled CL at different diameters using beam spot illumination at 0.5 keV ($\delta\approx 10$ nm~\cite{raftari2018modified}) and 2 keV ($\delta\approx 175$ nm~\cite{raftari2018modified}). 
Results of these measurements are shown in Fig.~\ref{fig:results3}. The PLDOS curve is calculated at $y=0$ for the respective values of $\delta$ given above.
The experimental results show generally good qualitative and quantitative agreement with the calculated PLDOS curve. In particular, the difference in the PLDOS peak position and the difference in the peak widths is clearly reproduced by the data. For the 0.5 keV data, we observe a peak at $s=1.4$ whereas for 2.0 keV the peak occurs at $s=1.9$. This corresponds to a difference in radius of $100$ nm. 

\section{Discussion}
In this work, we defined the PLDOS for the fundamental mode of an optical fiber and experimentally evalutated the PLDOS by
measuring CL coupled directly to the fiber fundamental modes. Using this technique, we made the first complete measurements of the PLDOS dependence on the size parameter around the diffration limit. We clearly demonstrated the different PLDOS behavior for points near the fiber surface and nearer to the fiber center. Although previous CL measurements of photonic crystal waveguide modes do exist, they have relied on intrinsic losses or leaky modes which coupled to the far field~\cite{brenny2016near}. Likewise, although the coupling efficiency from point emitters to the modes of a fiber has been measured, these measurements suffered from large systematic errors and did not reveal the full behavior of the PLDOS itself~\cite{PhysRevLett.109.063602}.  In contrast we are able to clearly measure the difference in PLDOS behavior near the fiber surface and nearer to the fiber center even though the respective PLDOS peak positions differ by a fiber radius of just 100 nm.

This work successfully enlarges the domain in which CL spectroscopy may be applied, from its original application to modes with a radiative component to the case of completely bound photonic states of which the modes of a waveguide are one example. It should also be possible to use our technique to couple electron beam induced luminescence from more general non-radiative modes which do not couple to the far field. Such modes can couple via the evanescent field of the optical fiber taper to its guided modes and thus be detected as in the present experiment, opening up CL spectroscopy to regimes which could traditionally only be measured using electron energy loss (EEL) methods. Due to the much less rigorous requirements for sample preparation and electron beam energy required for CL spectroscopy as compared with EEL spectroscopy, this is a significant addition to the electron spectroscopy toolbox.

In terms of applications typical fiber coupled photon sources up to now have used optically excited emitters~\cite{fujiwara2011highly,PhysRevLett.109.063602, yalla2014cavity}. Our method should provide a new route to achieving waveguide-coupled, electrically driven photon sources~\cite{le2013electrically,tizei2013spatially,meuret2015photon}. In particular, the ability to simultaneously image the nanostructure surface and excite fiber coupled cathode luminescence will allow a more deterministic approach even for non-deterministically assembled composite nanodevices created by combining nanowaveguides with colloidal nanocrystals.

For the above reasons, we believe that the technique detailed here  can open new opportunities to study fundamental aspects of nano-optics by measuring PLDOS through waveguide modes, while also providing a new platform for applications.

This work was supported by the Nano-Quantum Information Research Division of Tokyo University of Science. Part of this work was supported by JST CREST (Grant Number JPMJCR18I5).

%
\appendix

\section{Fiber guided modes}
Treatments of the guided modes of step-index optical fibers may be found in a number of places~\cite{okamoto2006fundamentals,le2004field}. 
For convenience, we present a treatment of the mode functions that follows references~\cite{le2005spontaneous,le2017higher}.

The wave equation in cylindrical coordinates for the $z$ component of an electromagnetic mode $\mathbf{E}(r,\phi)$ propagating along the $z$-axis with radial coordinate $r$ and azimuthal coordinate $\phi$ is
\begin{equation}
\frac{\partial^2 E_z}{\partial r^2} + \frac{1}{r}\frac{\partial E_z}{\partial r} + \frac{1}{r^2}\frac{\partial^2 E_z}{\partial \phi^2} + [k^2n^2 - \beta^2]E_z = 0,
\label{eq:waveq1}
\end{equation}
where $k=2\pi/\lambda$ is the free space wave number, $n=n(r)$ is the refractive index, and $\beta$ is the mode propagation constant. Setting $E(r,\phi)=e(r)e_\phi(\phi)$, and taking $e_\phi(\phi)=\exp({\rm i}m\phi)$ (requiring integer $m$), the radial wave equation is found to be
\begin{equation}
\frac{\partial^2 e_z}{\partial r^2} + \frac{1}{r}\frac{\partial e_z}{\partial r} + \left[\chi^2-\frac{m^2}{r^2}\right]e_z = 0,
\label{eq:waveq2}
\end{equation}
where $\chi^2 = k^2n^2-\beta^2$. Specializing to a step index fiber of radius $a$ where the core index is $n_{\rm co}$ and the cladding index is $n_{\rm cl}$, we split $\chi^2$ into two cases: $h^2=k^2n_{\rm co}^2-\beta^2$ in the core, and $q^2=\beta^2 - k^2n_{\rm cl}^2$ in the cladding. Full consideration of boundary conditions restricts the solutions to 
\begin{equation}
e_z = A\frac{2q}{\beta}\frac{K_m(qa)}{J_m(qa)}J_m(qr), \; r\leq a,
\end{equation}
and
\begin{equation}
e_z = A\frac{2q}{\beta} K_m(qr), \; r>a,
\end{equation}
for an arbitrary amplitude $A$. It can be shown that the radial and azimuthal components can be derived from $e_z$.
$J_m$ and $K_m$ are Bessel functions of the first kind and modified Bessel functions of the second kind respectively, with order $m$.

Restricting ourselves to the fundamental mode with $m=1$, and taking a clockwise circular polarization, the mode function components are
\begin{eqnarray}
 e_r& =& {\rm i}A\frac{q}{h}\frac{K_1(qa)}{J_1(qa)}[(1-s)J_0(hr) - (1+s)J_2(hr)]\nonumber\\
 e_\phi& =& -A\frac{q}{h}\frac{K_1(qa)}{J_1(qa)}[(1-s)J_0(hr) - (1+s)J_2(hr)] \nonumber\\
 e_z &=& A\frac{2q}{\beta}\frac{K_1(qa)}{J_1(qa)}J_1(qr) \nonumber 
\end{eqnarray}
in the core and 
\begin{eqnarray}
 e_r& =& {\rm i}A[(1-s)K_0(hr) - (1+s)K_2(hr)]\nonumber\\
 e_\phi& =& -A[(1-s)K_0(hr) - (1+s)K_2(hr)] \nonumber\\
 e_z &=& A\frac{2q}{\beta}K_1(qr) \nonumber 
\end{eqnarray}
in the cladding. Here, we have $s = (1/q^2a^2 + 1/h^2a^2) / (J'_1(ha)/haJ_1(ha) + K'_1(qa)/qaK_1(qa))$.

To produce the mode functions, we choose $A$ so that $\int d^2 r n(r)^2|\mathbf{e}|^2$=1, where the integral is taken over the entire $r-\phi$ plane. For brevity, we omit the expression for the integral, along with the eigenvalue equation required to find $\beta$. The appropriate expressions may be found elsewhere~\cite{le2005spontaneous,le2017higher}. We note that the left hand side of the normalization condition is related to but not identical to the mode power.

Inside the fiber core (as is the case in the current work) we find 
\begin{eqnarray}
|\mathbf{e}|^2 &=& |e_r|^2 + |e_\phi|^2 + |e_z|^2\nonumber\\ 
 &=& 2A^2\frac{q^2K_1^2(qa)}{h^2J_1^2(ha)}\left[(1-s)^2J_0^2(hr) + \frac{h^2}{\beta^2}J_1^2(hr) \right.+ \nonumber\\
 & & \left. (1+s)^2J_2^2(hr)\right].\nonumber
\end{eqnarray}

In order to make clearer the contributions to the PLDOS, we divide the mode function intensity into $r$ independent and dependent parts as follows:
\begin{equation}
|\mathbf{e}|^2 = A^2(k,a)F^2(k,a,r),
\end{equation}
where 
\[
A^2(k,a) = A^2\frac{q^2K_1^2(qa)}{h^2J_1^2(ha)}
\]
and 
\[
F(k,a,r) = (1-s)^2J_0^2(ur/a) + \frac{h^2}{\beta^2}J_1^2(ur/a) + (1+s)^2J_2^2(ur/a),
\]
where $u=ah$.

\begin{figure}
\centering
\includegraphics[width=0.8\linewidth]{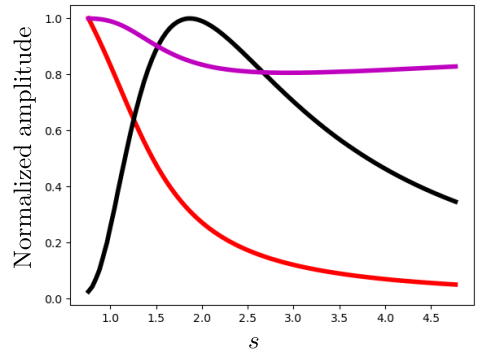}
\caption{\label{fig:AF} $A(k,a)$ (black curve), $F(k,a,r=0)$ (magenta curve) and$F(k,a,r=a)$ (red curve) .}
\end{figure}

From Fig.~\ref{fig:AF}, it may be seen that $A(k,a)$ (black curve) has a peaked form and is responsible for the overall shape of the PLDOS,
as discussed in the main text. $F(k,a,r)$ for a set value of $r/a$ is a decaying function of the size parameter $s$, with the decay rate being smaller at the fiber center ($r=0$, magenta line in Fig.~\ref{fig:AF} ) than at the fiber surface ($r=a$, red line in Fig.~\ref{fig:AF}). When multiplied by $A(k,a)$, this behavior of the $F$ function explains both the shift in the PLDOS peak depending on $r$ and the width of the PLDOS peak.

\section{Details of the experiment}
\begin{figure*}
\centering
\includegraphics[width=\linewidth]{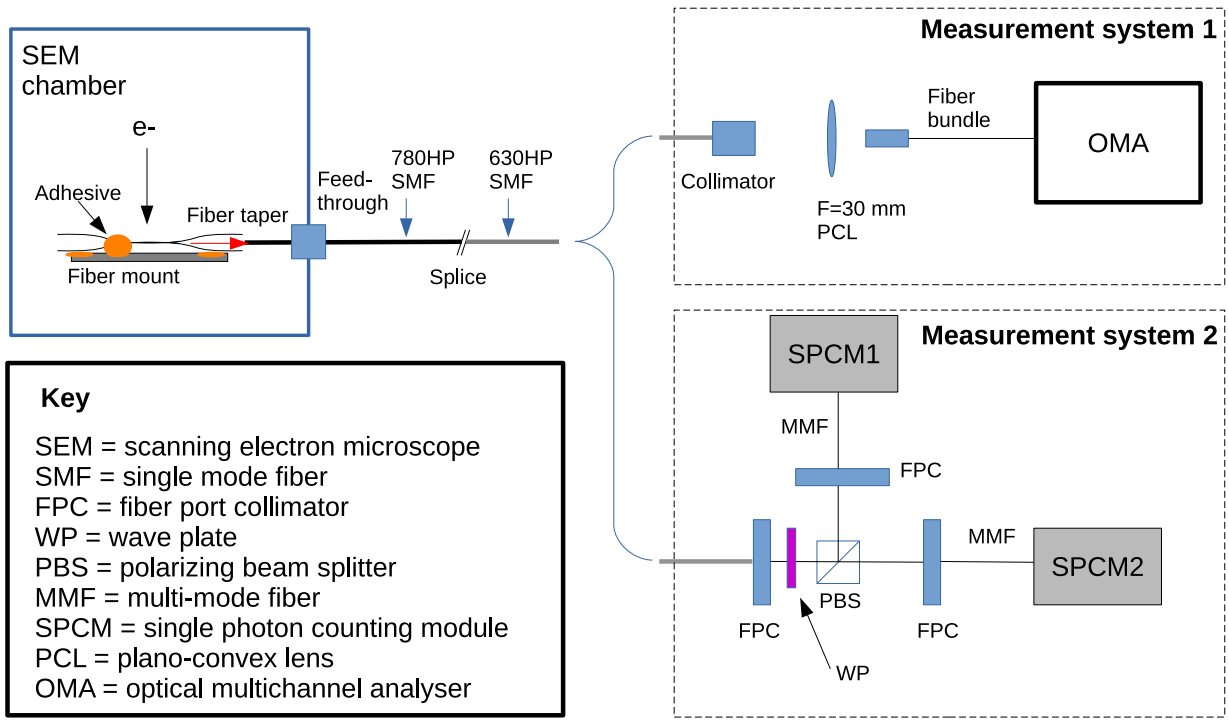}
\caption{\label{fig:setup}Experimental setup. Electrons produced by SEM gun are focussed and incident on an tapered, vacuum clad optical fiber which is mounted in the SEM vacuum chamber. The optical fiber tapers adiabatically into a standard optical fiber which passes through a feedthrough and can be connected to one of two measurement systems. Measurement system 1 allows the measurement of the CL spectrum. Measurement system 2 allows the measurement of CL intensity, polarization and the correlation of CL photons. Acronyms used are explained in the Key. }
\end{figure*}
The experimental setup is depicted schematically in Fig.~\ref{fig:setup}.
We used the electron beam of a scanning electron microscope (LEO 1530VP, Carl Zeiss) to
excite CL in our sample. The sample chamber was evacuated with a turbo-molecular pump down to $1 \times 10^{-3}\ \mathrm{Pa}$. The primary-electron column is a Gemini type which achieves high resolution for low energy electrons compared to a conventional SEM~\cite{jaksch1995high}. A schottky field emission electron source (SFE) is installed in the SEM gun chamber. The SFE has a very low beam noise and notable long term beam current stability. Primary SEM observations were made in an electron energy range of $0.5 - 2.0\ \mathrm{keV}$. The beam current was measured using a Faraday cup yielding approimately 40 pA. The electron beam profile was evaluated using Au-Pd coated polystyrene latex spheres, of $90\ \mathrm{nm}$ in diameter~\cite{irita2018development,irita2018compact}. 
The spatial resolution (20/80$\%$ edge profile) was about 5 nm in the electron energy range used in the experiment. The electron beam was used to excite luminescence in an optical fiber taper (see below) using either a stationary spot excitation mode, or a sweep excitation mode, where the electron beam was scanned over the fiber, allowing imaging by detection of secondary electrons. 

Regarding the optical setup,
the tapered fiber was manufactured from a commercial single mode fiber (780 HP) using a heat and pull technique~\cite{Sile}.
Tapered fibers used in the experiment had a transmission of at least $90\%$ and a typical transmission of $95\%$.
The fiber was mounted in the SEM and its output was spliced to a standard optical fiber which passed out of the SEM through a homemade feedthrough system~\cite{Abraham:98}. Regarding the mounting of the fiber taper: we used a UV cured adhesive to fix the fiber to an aluminium mount at two points maximally far from the taper center. To suppress vibrations of the fiber, we also added adhesive to one side of the taper closer to the taper center, meaning that fluorescence could only be measured through one of the fiber outputs, due to strong absorption and scattering caused by the adhesive. We note that CL can still be induced in the event of fiber vibrations, but precise measurement of the fiber diameter, as required for the current experiment, is difficult.

For CL spectrum observation, the output fiber was connected to a spectrometer (ACTON Spectra Pro 2300, Princeton Instruments) equipped with a CCD detector (Pixis 100BR, Princeton Instruments) to measure the wavelength as depicted by Fig.~\ref{fig:setup}, Measurement System 1.  In order to measure the intensity of CL, photon polarization, and photon correlations, we used Measurement System 2 as shown in Fig.~\ref{fig:setup}. We used a fiber u-bench setup with a polarizing beam splitter installed whose outputs were coupled to multimode fibers which were in turn connected to single photon counting modules (SPCM-AQRH-14-FC, Excelitas). Count rates and photon correlation measurements were made using a two channel counter / correlator (TimeTagger20, Swabian Instruments).

Note that in all optical detection experiments, we spliced the output of the main fiber (780HP), single mode above 780 nm in wavelength) to a fiber which was single-mode at our operating wavelength (630HP) in order to guarantee that we only measured light coupled to the fundamental mode of the fiber.

\bibliography{SEMNanofiber}

\end{document}